\documentclass{article}

\usepackage{arxiv}

\usepackage[utf8]{inputenc} 
\usepackage[T1]{fontenc}    

\usepackage{booktabs}       
\usepackage{amsfonts}       
\usepackage{nicefrac}       
\usepackage{microtype}      
\usepackage{svg}
\usepackage{url}            

\title{AVECL-UMONS database for audio-visual event classification and localization}

\author{
  Mathilde Brousmiche\\
  Numediart Institute\\
  University of Mons, Belgium\\
  \texttt{mathilde.brousmiche@umons.ac.be} \\
   \And
  Stéphane Dupont\\
  Numediart Institute\\
  University of Mons, Belgium\\
  \texttt{stephane.dupont@umons.ac.be} \\
  \AND
  Jean Rouat\\
  NECOTIS Lab\\
  University of Sherbrooke, Canada\\
  \texttt{jean.rouat@usherbrooke.ca} \\
}

\begin{document}
\maketitle

\begin{abstract}
We introduce the AVECL-UMons dataset for audio-visual event classification and localization in the context of office environments.   The audio-visual dataset is composed of  11  event classes recorded at several realistic positions in two different rooms. Two types of sequences are recorded according to the number of events in the sequence.  The dataset comprises 2662 unilabel sequences and 2724 multilabel sequences corresponding to a total of 5.24 hours. The dataset is publicly accessible online : \url{https://zenodo.org/record/3965492#.X09wsobgrCI}.
\end{abstract}


\section{Introduction}

In computer vision, different datasets were created for event classification based on visual information such as YouTube-8M \cite{real2017youtube}, Kinetics \cite{carreira2017quo} and Sports-1M \cite{karpathy2014large} UCF101 \cite{soomro2012ucf101}, etc. In recent years, researchers have decided not only to exploit visual information, but also to exploit sound information included in video. As the video datasets do not always include relevant sound information, Tian \textit{et al.} create the AVE dataset for audio-visual event detection \cite{tian2018audio}.

Also in the context of computer vision, other datasets were created to localize different objects in the image either with bounding box (PASCAL VOC \cite{everingham2015pascal}, Objects365 \cite{shao2019objects365}, etc.) or with segmentation at the pixel level (COCO \cite{lin2014microsoft}, Cityscapes \cite{cordts2016cityscapes}, CamVid \cite{brostow2009semantic}, etc.). Localization in videos is rarer, only a few datasets exist \cite{real2017youtube, gu2018ava}. The localization in the image is interesting information but for indoor events, it may be interested to locate the event in the room.

In the other hand, in computer audition, there is the Sound Event Localization and Detection task (SELD). SELD can be divided into two subtasks: Sound Event Detection (SED) and Sound Source Localization (SSL). The SELD goal includes recognizing each sound event class present in the acoustic scene and simultaneously locate in space each detected sound event.

Several datasets were created for Sound Event Detection and Classification (Urban Sound \cite{salamon2014dataset}, TUT Sound events \cite{mesaros2016tut}, etc.) or for Sound Source Localization ((MUSLOD) \cite{meraoubi2014multimicrophone}, AV16.3 \cite{lathoud2004av16}, etc.) but very few for Sound Event Detection and Localization (multi-room reverberant dataset \cite{adavanne2019multi}). Furthermore, the multi-room reverberant dataset is composed of simulated data, meaning that real-life impulse responses are convolved with isolated sound events from the DCASE 2016 task to create the dataset. All these datasets are only composed of sound and do not comprise the visual information.

We recorded a new dataset of audio-visual events in the context of office environments. The dataset consists of two sets : SECL-UMONS includes recordings of different events with a microphone array and AVECL-UMONS includes recordings of the same events with four webcams. Both sets are available on Zenodo.

This report contains information about webcam recordings (AVECL-UMONS). Information on microphone recordings (SECL-UMONS) can be found in \cite{brousmiche2020secl}.

\section{Data collection}
\label{sec:data_collection}

The dataset is divided into two parts according to the number of events in the sequence: unilabel sequences (one event per sequence) or multilabel sequences (two simultaneous events per sequence).

\subsection{Recording conditions}

\begin{figure}[h]
    \begin{minipage}[b]{.48\linewidth}
      \centering
      \centerline{\includegraphics[width=6.0cm]{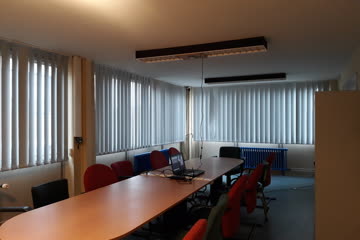}}
      \centerline{(a) Room 1 : RT60 of 0.7 s}
      \centerline{Dim [m] : 7.8 x 3.6 x 2.45}
      \medskip
    \end{minipage}
    \hfill
    \begin{minipage}[b]{0.48\linewidth}
      \centering
      \centerline{\includegraphics[width=6.0cm]{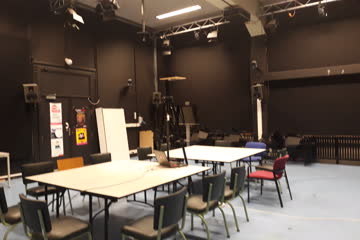}}
      \centerline{(b) Room 2 : RT60 of 0.9 s}
      \centerline{Dim [m] : 9.40 x 7.5 x 4.85}
      \medskip
    \end{minipage}
    
    \caption{Visualization of the two rooms used for the recordings}
    \label{fig:rooms}
\end{figure}

\begin{figure}[h]
     \begin{minipage}[b]{.48\linewidth}
      \centering
      \centerline{\includegraphics[width=9cm]{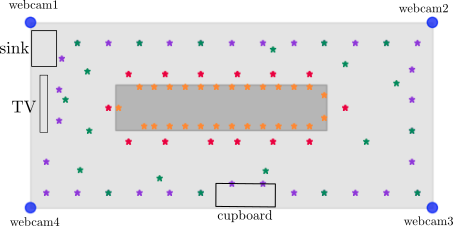}}
      \centerline{(a) Room 1}
      \medskip
    \end{minipage}
    \hfill
    \begin{minipage}[b]{0.48\linewidth}
      \centering
      \centerline{\includegraphics[width=8.5cm]{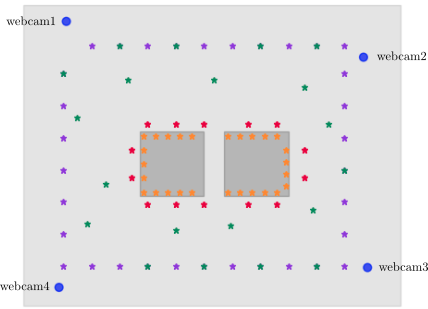}}
      \centerline{(b) Room 2}
      \medskip
    \end{minipage}
    \caption{Diagram of Room 1. The other dots are the possible positions in the room for the different event classes. Orange : Cup drop off, Keyboard, Phone ring; Red : Chair movement, Hand Clap, Speaker, Whistle; Green : Hand Clap, Speaker, Step, Whistle; Purple: Furniture's drawer, Knock, Step.}
    \label{fig:room1}
\end{figure}

Audio-visual events are recorded with 4 Logitech C920 webcams \footnote{\url{https://www.logitech.com/en-us/product/hd-pro-webcam-c920}}. We used OBS software with 20 frame per second and frame dimension of $1920 \times 1080$. The webcams includes stereo microphones (sampling rate of $44100$ Hz). Events are recorded in two different rooms (Fig. \ref{fig:rooms}). Depending on the event classes, different positions are possible in the room (Fig. \ref{fig:room1}).

A microphone array was also used, for more details about the multichannel sound dataset see \cite{brousmiche2020secl}.

\subsection{Unilabel sequences}

The dataset comprises 11 classes having several subclasses. The difference between subclasses is either the use of a different object belonging to the same class or a different participant performing the action. For each class, several positions are marked in the two rooms before starting the recordings.

Recordings of several minutes, named session, are saved. Afterward, the sequences of interest containing only the audio-visual event are extracted. One session is realized for each subclass. During a session, the participant realizes the event at each possible position marked beforehand. A script, shown on a screen in the room, is run for each session. When and where the events have to occur are ordered by this script. To avoid the presence of the noise of the participant movement, when to move between two events is also ordered by the script. Afterward, sequences of interest are automatically extracted from the session recordings thanks to the time planned by the script. A total of 2662 sequences of 3 seconds composed of only one event are extracted from the different session recordings.

\begin{figure}[h]
    \centering
    \includegraphics[width=10cm]{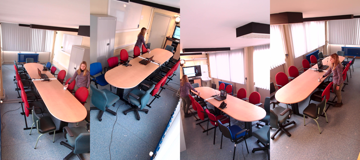}
    \caption{Example of unilabel data for each webcam in Room 1.}
    \label{fig:unilabel}
\end{figure}

\subsection{Multilabel sequences}

Multilabel sequences are composed of two event classes realized approximately at the same time. The combinations of two event classes are chosen for all possible duos of events. Therefore, each event class (except \textit{furniture's drawer}) is associated with all classes, even the same class. Therefore, there are 55 possible associations. A session is realized for each duo of event classes. Both participants take different positions in the room. 25 positions are selected to cover a maximum of situations (all around the room, away from each other and close from each other). Again, a script is run for each session. When and where the events have to occur as well as when to move to avoid parasitic noise is ordered by this script. Afterward, sequences are extracted from the session recordings thanks to the time planned with the script. A total of 2724 sequences of 4 seconds composed of two events are extracted.

\begin{figure}[h]
    \centering
    \includegraphics[width=10cm]{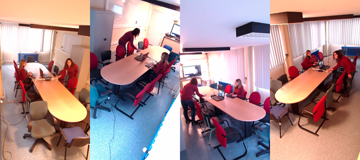}
    \caption{Example of multilabel data for each webcam in Room 1.}
    \label{fig:multilabel}
\end{figure}

\subsection{Metadata}

As metadata, we provide different information. 

For each unilabel sequence, the event class, the subclass, the x,y,z coordinates in the room, the number of the room are provided. For each webcam, the presence or not of the event in the field of view is provided.

For multilabel sequences, the same information is provided but are provided for the first and the second class.

\section*{Acknowledgement}
Thanks to CHISTERA IGLU and the European Regional Development Fund (ERDF) for funding.

\bibliographystyle{unsrt}  
\bibliography{references}  
\end{document}